\documentclass[twocolumn,apjfonts,tighten]{aastex631}

\usepackage{apjfonts}
\usepackage{graphicx}
\usepackage{wrapfig}
\usepackage{amsmath}
\usepackage{multirow}
\usepackage{booktabs}
\usepackage{float}
\usepackage{gensymb}

\newcommand{\Msun}{$M_{\sun}$}

\defcitealias{weisz_high-mass_2015}{W15}

\shorttitle{PHATTER. VI. The High-Mass Stellar IMF in M33}
\shortauthors{Wainer et al.}

\begin{document}

\title{The Panchromatic Hubble Andromeda Treasury: Triangulum Extended Region (PHATTER). \\
VI. The High-Mass Stellar Initial Mass Function of M33}

\author[0000-0001-6320-2230]{Tobin M. Wainer}
\affiliation{Department of Astronomy, University of Washington, Box 351580, Seattle, WA 98195, USA}

\author[0000-0002-7502-0597]{Benjamin F. Williams}
\affiliation{Department of Astronomy, University of Washington, Box 351580, Seattle, WA 98195, USA}

\author[0000-0001-6421-0953]{L. Clifton Johnson}
\affiliation{Center for Interdisciplinary Exploration and Research in Astrophysics (CIERA) and Department of Physics and Astronomy, Northwestern University, 1800 Sherman Ave., Evanston, IL 60201, USA}

\author[0000-0002-6442-6030]{Daniel R. Weisz}
\affiliation{Department of Astronomy, University of California Berkeley, Berkeley, CA 94720, USA}

\author[0000-0002-1264-2006]{Julianne J. Dalcanton}
\affiliation{Department of Astronomy, University of Washington, Box 351580, Seattle, WA 98195, USA}
\affiliation{Center for Computational Astrophysics, Flatiron Institute, 162 Fifth Avenue, New York, NY 10010, USA}

\author[0000-0003-0248-5470]{Anil C. Seth}
\affiliation{Department of Physics and Astronomy, University of Utah, Salt Lake City, UT 84112, USA}

\author[0000-0001-8416-4093]{Andrew Dolphin}
\affiliation{Raytheon, Tucson, AZ 85756, USA}
\affiliation{Steward Observatory, University of Arizona, Tucson, AZ 85726, USA}

\author[0000-0001-7531-9815]{Meredith J. Durbin}
\affiliation{Department of Astronomy, University of California Berkeley, Berkeley, CA 94720, USA}
\affiliation{Department of Astronomy, University of Washington, Box 351580, Seattle, WA 98195, USA}

\author[0000-0002-5564-9873]{Eric F. Bell}
\affiliation{2 Department of Astronomy, University of Michigan, Ann Arbor, MI, USA}

\author[0000-0002-3038-3896]{Zhuo Chen}
\affiliation{Department of Astronomy, University of Washington, Box 351580, Seattle, WA 98195, USA}

\author[0000-0001-8867-4234]{Puragra Guhathakurta}
\affiliation{Department of Astronomy and Astrophysics, University of California Santa Cruz, University of California Observatories, 1156 High Street, Santa Cruz, CA 95064, USA}

\author[0000-0001-9605-780X]{Eric W. Koch}
\affiliation{Center for Astrophysics $\mid$ Harvard \& Smithsonian, 60 Garden St., 02138 Cambridge, MA, USA}

\author[0000-0003-0588-7360]{Christina W. Lindberg}
\affiliation{Johns Hopkins University, 3400 North Charles St., 473 Bloomberg Center for Physics and Astronomy, Baltimore, MD, 21218}

\author[0000-0002-5204-2259]{Erik~Rosolowsky}
\affiliation{Department of Physics, University of Alberta, Edmonton, AB T6G 2E1, Canada}

\author[0000-0002-4378-8534]{Karin M. Sandstrom}
\affiliation{Department of Astronomy \& Astrophysics, University of California, San Diego, 9500 Gilman Drive, La Jolla, CA 92093, USA}

\author[0000-0003-0605-8732]{Evan D.\ Skillman}
\affiliation{Minnesota Institute for Astrophysics, University of Minnesota, Minneapolis, MN 55455, USA}

\author[0000-0003-2599-7524]{Adam Smercina}
\affiliation{Department of Astronomy, University of Washington, Box 351580, Seattle, WA 98195, USA}

\author[0000-0001-9961-8203]{Estephani E. TorresVillanueva}
\affiliation{Department of Astronomy, University of Wisconsin-Madison, Madison, WI, 53706, USA}

\correspondingauthor{Tobin M. Wainer}
\email{tobinw@uw.edu}

\begin{abstract}
We measure the high-mass stellar initial mass function (IMF) from resolved stars in M33 young stellar clusters. Leveraging \textit{Hubble Space Telescope's} high resolving power, we fully model the IMF probabilistically. We first model the optical CMD of each cluster to constrain its power-law slope $\Gamma$, marginalized over other cluster parameters in the fit (e.g., cluster age, mass, and radius). We then probabilistically model the distribution of MF slopes for a highly strict cluster sample of 9 clusters more massive than log(Mass/M$_{\odot}$)=3.6; above this mass, all clusters have well-populated main sequences of massive stars and should have accurate recovery of their MF slopes, based on extensive tests with artificial clusters. We find the ensemble IMF is best described by a mean high-mass slope of $\overline{\Gamma} = 1.49\pm0.18$, with an intrinsic scatter of $\sigma^{2}_{\Gamma} = 0.02^{+0.16}_{0.00}$, consistent with a universal IMF. We find no dependence of the IMF on environmental impacts such as the local star formation rate or galactocentric radius within M33, which serves as a proxy for metallicity. This $\overline{\Gamma}$ measurement is consistent with similar measurements in M31, despite M33 having a much higher star formation rate intensity. While this measurement is formally consistent with the canonical Kroupa ($\Gamma = 1.30$) IMF, as well as the Salpeter ($\Gamma = 1.35)$) value, it is the second Local Group cluster sample to show evidence for a somewhat steeper high-mass IMF slope. We explore the impacts a steeper IMF slope has on a number of astronomical sub-fields.
\end{abstract}


\section{Introduction}
The stellar initial mass function (IMF) plays a central role in both observational and theoretical astrophysics. Formally, the IMF is the distribution of the masses of stars formed in a single event. Of particular interest is the stellar IMF for stars greater than 1 M$_{\odot}$. Because such stars evolve over cosmologically short timescales to become supernovae and asymptotic giant branch stars that energize and enrich galaxies, their relative numbers affect a wide range of astrophysics. 

Observationally, the high-mass IMF impacts interpretations of stellar populations of galaxies \citep[e.g.,][]{bell_optical_2003, arrabal_haro_confirmation_2023}, cosmic reionization \citep[e.g.,][]{ouchi_large_2009}, and star formation efficiency \citep[e.g.,][]{alves_mass_2007, hennebelle_analytical_2008}. Theoretically, the IMF dictates prescriptions in numerical simulations of galaxy formation \citep[e.g.,][]{kim_agora_2014}, chemical yields in simulations \cite[e.g.,][]{buck_challenge_2021}, modeling the frequency of core-collapse supernovae \citep[e.g.,][]{heger_how_2003}, and constraining stellar evolution models \citep[e.g.,][]{massey_initial_1998, kroupa_variation_2001, krumholz_big_2014}. 

The formation of high-mass stars has also been shown to affect the overall star forming process for other stars in the same region \citep[e.g.,][]{grudic_does_2023}. Specifically, the formation of a massive star early or late in the star formation process can have important impacts on the overall star formation in a molecular cloud \citep{dinnbier_dynamical_2022, lewis_early-forming_2023}. Thus, quantifying the number of high mass stars created is vital for a comprehensive understanding of the entire star formation process.

Given these widespread impacts, the high-mass IMF has long been of great interest \citep[see review by][]{bastian_universal_2010}. Most recently, theoretical studies of the IMF have attempted to quantify the impact that different physical and environmental effects, such as metallicity and star formation rate, have on the high-mass IMF \citep[e.g.,][]{li_stellar_2023, grudic_great_2023, grudic_does_2023, smith_can_2023, parravano_high-mass_2018, zhang_stellar_2018, marks_evidence_2012}. These studies have further increased the motivation for observational measurements of the high-mass IMF across a range of environments.

Historically, the IMF was first measured in the 1950's by counting stars in the solar neighborhood \citep{salpeter_luminosity_1955}, and modeled with a two part power law ($dN / dM \propto M^{\Gamma_i}$) where $\Gamma_i$ denotes the slopes of the two power laws, with a break at 1 M$_{\odot}$. Early studies focused mainly on the universality of the IMF \citep[e.g.,][]{lequeux_quantitative_1979, miller_initial_1979} and whether it was possible to observe any considerable variations from the canonical \citet{salpeter_luminosity_1955} IMF. This led some to argue that there does not appear to be a systematic IMF dependence on environment or initial star-forming conditions \citep[e.g][]{bastian_universal_2010, offner_origin_2014}, although other studies present evidence for a top-heavy IMF in the galactic center \citep[e.g.,][]{lu_stellar_2013}. These studies, however, did not explain the observed scatter in measured IMF slopes, contradicting a picture where the IMF was universal across environments. However, \citet{weisz_panchromatic_2013} showed that $>75\%$ of IMF studies significantly underestimated uncertainties, detailing how difficult it is to untangle systematic effects from intrinsic IMF variability.

Due to these limitations, our understanding of the high-mass IMF is incomplete. In spite of the uncertain observational evidence, within the field there has been an adoption of a ``universal" IMF for theoretical predictions. This typically assumes either the Kroupa \cite[$\Gamma = +1.3$;][]{kroupa_variation_2001} or Salpeter \citep[$\Gamma = +1.35$;][]{salpeter_luminosity_1955} slope for stars more massive than 1 M$_{\odot}$, assuming no variations, despite persistent controversy over the universality of the IMF since the first studies \citep[e.g.,][]{scalo_stellar_1986, scalo_imf_1998, kroupa_initial_2002}.

Current observational approaches to measure the high-mass IMF and its applicability fall primarily into two large classes. The first approach uses stellar systems where individual stars are not resolvable. Studies have characterized galactic-wide stellar mass functions for unresolved sources through integrated light techniques \citep[e.g.,][]{lee_comparison_2009, meurer_evidence_2009, conroy_stellar_2012, dib_galaxy-wide_2022, cheng_initial_2023}. These studies require substantial assumptions due to degeneracies between dust, age, metallicity, and stochastic sampling of the IMF itself. As such, it remains undetermined whether integrated light measurements will provide a new avenue for reliably characterizing the IMF \citep[e.g.,][]{calzetti_method_2010, weidner_sampling_2014, weisz_panchromatic_2013}.  

In contrast to "unresolved" approaches to measure the high-mass IMF, our current best constraints come from studying resolved stellar populations in the Milky Way and nearby galaxies. One notable study is that of \citet[][hereafter W15]{weisz_high-mass_2015}, who measured the IMF for 85 star clusters in the north-east third of M31 using data from the Panchromatic Hubble Andromeda Treasury (PHAT) data \citep{dalcanton_panchromatic_2012}. This study remains one of the best extragalactic IMF measurements, yielding one of the most robust high-mass IMF measurements to date. They found a universal high-mass IMF unlikely to be the best representation in the Local Group. Instead they find a slightly steeper high-mass IMF where $\Gamma = 1.45^{+0.03}_{-0.06}$. Although similar to the traditional values of the \citeauthor{salpeter_luminosity_1955} IMF, the deviation makes a considerable difference in the number of high mass stars formed and in the values obtained from common star formation rate prescriptions.

The tension in the debate between a perfectly known, universal IMF, and an IMF dependent on environment and chemical properties, remains high, as discussed in the recent review article by \citet{lee_origin_2020} \citep[and complementary articles, e.g.,][]{escher_jung_universal_2023, baumgardt_evidence_2023, pouteau_alma-imf_2022, zhang_stellar_2018, marks_evidence_2012}. Deviations from a universal IMF could manifest in both bottom-heavy \citep[e.g.,][]{spiniello_evidence_2012, smith_evidence_2020} IMFs of low mass stars ($<$1 M$_{\odot}$), and top-heavy \citep[e.g.,][]{marks_evidence_2012, pouteau_alma-imf_2022} IMFs of massive stars discussed in this work. Due to the observed cluster-to-cluster variance of the IMF, a large, homogeneous survey of individually resolved young, massive stars is required to reliably resolve these debates \citepalias{weisz_high-mass_2015}. Thus, the need for additional IMF studies in galaxies where we can resolve individual stars is required for our ability to understand if there is a universal high-mass IMF. 

One of the best targets for such a study is the Local Group spiral M33. M33 is a nearby galaxy to us with a relatively face on inclination of $55^{\degree}$ \citep{koch_kinematics_2018}, and also has lower line-of-sight dust attenuation compared to M31. With a distance modulus of 24.67 \citep{de_grijs_clustering_2014}, it is possible to resolve individual stars down to at least $1M_{\odot}$. However, compared to M31, M33 has a lower metallicity, and a higher SFR intensity, making it a natural target for identifying any environmentally driven variations in the high-mass IMF. 

In M33, the high-mass IMF has been studied through inferences of integrated properties \citep[e.g.,][]{grossi_young_2010, corbelli_cluster_2011}, as well as single cluster IMF characterizations \citep[e.g.,][]{gonzalez_delgado_massive_2000, jamet_importance_2004}. However, the high-mass IMF has not previously been measured across star clusters with resolved stars, the most reliable characterization. We follow the methodology of \citetalias{weisz_high-mass_2015}, measuring the IMF from resolved stars above $1M_{\odot}$ probabilistically. We model the optical CMD of each cluster to constrain its power law slope, $\Gamma$, of the present-day high-mass MF, marginalized over other cluster parameters (i.e., cluster age, mass, and radius).

This paper is structured as follows. In Section~\ref{sec: data}, we describe the PHATTER photometry, define our cluster sample, and describe our synthetic cluster sample. In Section~\ref{sec: analysis}, we lay out the probabilistic approach for MF fitting and the inference of the IMF from these measurements. We present our results in Section~\ref{sec: results} and discuss implications in Section~\ref{sec: discusion}.

\section{Data}\label{sec: data} 
Our cluster data is drawn from the Local Group Cluster Search (LGCS) cluster catalog \citep{johnson_panchromatic_2022}, which was created using data from the PHATTER survey \citep{williams_panchromatic_2021}. The PHATTER survey uses the \textit{Hubble Space Telescope} (HST) Advanced Camera for Surveys (ACS) and Wide Field Camera 3 (WFC3) to image the inner 5 kpc of M33's disk in six filters from UV to IR. The \citet{johnson_panchromatic_2022} catalog presents 1214 star clusters identified through a crowd-sourced, visual search of the optical PHATTER data (F475W \& F814W), facilitated by the LGCS project\footnote{\url{https://www.clustersearch.org}}, a citizen science effort hosted on the Zooniverse\footnote{\url{https://www.zooniverse.org}} platform. We adopt the recommended cluster catalog threshold (the minimum weighted fraction of identifications as a cluster, $f_{\textrm{cluster, W}} > 0.674$), which limits the contamination rate to $\lesssim$4\% for the cluster sample. Cluster properties for this sample were derived through CMD modeling in \citet{wainer_panchromatic_2022}.

\begin{figure*}[ht]
    \centering
    \includegraphics[width=.44\textwidth]{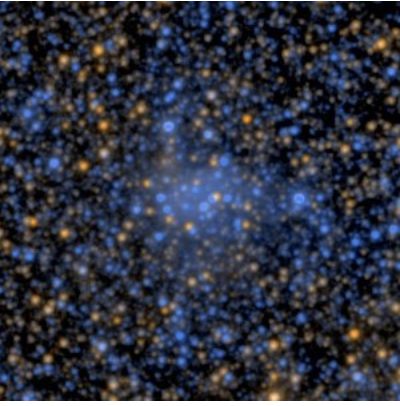}
    \includegraphics[width=.465\textwidth]{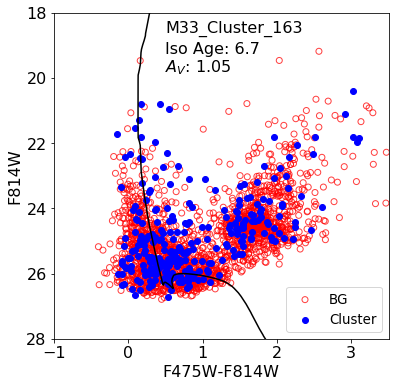}
    \caption{On the left is a F475W, F814W combined color image for M33 cluster 163 identified by \citet{johnson_panchromatic_2022} clearly showing resolved stars. On the right is the best fit isochrone to stellar photometry from \citet{wainer_panchromatic_2022}, highlighting the level of photometric depth available in the CMD. The blue dots are stars that fall within the photometric aperture, while the open red circles are stars within the background annulus (see text for details), illustrating the input data to MATCH.}
    \label{fig:img_cmd}
\end{figure*}

For this study, we follow the methodology of \citetalias{weisz_high-mass_2015} and select an initial, inclusive sample of clusters younger than 25 Myr, more massive than $10^3$M$_{\odot}$, and having half-light radii larger than 1.66 pc. These threshold limits were explicitly adopted due to the synthetic cluster analysis presented in Sections~\ref{sec:synthetics}, and represent the range of parameters where we can best recover the input IMF, which will be discussed further in Sections~\ref{sec:synthetics}. These clusters have ages of order $\sim$0.1 relaxation times, largely mitigating the effects of mass segregation or cluster dissolution, because these effects take place on longer time scales, making them the ideal sample to probe the high mass IMF \citepalias{weisz_high-mass_2015}. The lower mass limit excludes clusters that will not have sufficient stars to add meaningful constraints to the IMF. Selecting clusters with larger half-light radii limits the impacts of stellar crowding in the densest clusters. 

The M33 cluster sample from these selection cuts contains 34 clusters\footnote{We note that this sample does not include a few popular M33 targets. Most notably, NGC~604 is classified as a emission region opposed to a cluster in \citet{johnson_panchromatic_2022}, and NGC~588 which is not within the PHATTER footprint.}, compared to the 85 clusters analyzed in \citetalias{weisz_high-mass_2015}. The largest reason for this difference is the much larger survey area of PHAT in M31 compared to PHATTER in M33. In addition, there are subtle differences in our cluster selection. We include clusters with half-light radii greater than 1.66 pc as opposed to the 2.0 pc threshold used in \citetalias{weisz_high-mass_2015}. This difference in selection criterion includes eight additional clusters which would be excluded if we impose a 2.0 pc threshold. Including these clusters increased our sample by $\sim30\%$, allowing for a more precise measurement. We have performed extensive tests with synthetic clusters, and used these to determine our optimal cluster sub-sample, which we discuss in detail in Section \ref{sec:synthetics}.

For the purposes of this work, we use only the optical (F475W and F814W) resolved stellar photometry photometry from PHATTER, which contain the most information about the IMF slope due to having the deepest photometry, reaching F475W $\sim28$ mag, sufficient to detect $1.3M_\odot$ stars as seen in Figure~\ref{fig:completeness}, which we will further discuss in Section~\ref{sec: analysis}. The photometry of the PHATTER clusters was performed with \texttt{dolphot} \citep{dolphin_wfpc2_2000, dolphin_dolphot_2016}, using the HST specific modules described in \citet{williams_panchromatic_2014}. An example of the photometry is shown in the optical color image of M33 cluster 163, the left hand panel of Figure~\ref{fig:img_cmd}. The right hand panel has the CMD and best fit isochrone determined by \citet{wainer_panchromatic_2022}. 

The uncertainty, bias, and completeness of the resolved stellar photometry in each cluster is thoroughly tested with artificial star tests (ASTs), where artificial stars with known parameters are injected into the data and then recovered. Following the principles in \citet{williams_panchromatic_2021}, the artificial stars are generated with the \texttt{fake} utility of MATCH \citep{dolphin_numerical_2002}. For each cluster, we insert and recover 50,000 ASTs to ensure accurate characterization of photometric completeness and noise, as was done in \citetalias{weisz_high-mass_2015}. Input positions for the ASTs are distributed following the light profile of the cluster, based on the \citet{johnson_panchromatic_2022} half-light radii and assuming a \citet{king_structure_1962} profile.

To quantify the SFR of the surrounding region associated with each cluster in our sample, we use the SF maps of \citet{lazzarini_panchromatic_2022} who fit color magnitude diagrams to resolved stars in 2005, $\sim100$ parsec regions. Based on the cluster location, we adopt the \citet{lazzarini_panchromatic_2022} SFR value between 0-100 Myr ago as the cluster's local SFR. We normalize the value given in \citet{lazzarini_panchromatic_2022} by the area in order to estimate the $\Sigma_\mathrm{SFR}$, a value more comparable to galactic samples in the literature. We also compute the angular galactic radius of each cluster assuming a common distance modulus. 

\section{Analysis} \label{sec: analysis}
Following \citetalias{weisz_high-mass_2015}, we first measure the stellar probability distribution function (PDF) of the present day MF for each young cluster. We then fit the ensemble of the PDFs for the MF slopes to infer the intrinsic IMF slope for the ensemble of clusters \footnote{Throughout the rest of this paper, we will use "IMF" to indicate the ensemble, while individual cluster fits will be referred to as "MF".}.  We discuss each of these steps in detail below.

\subsection{Measuring Each Cluster's Stellar Mass Function}
\label{sec:Step1_MF_slope}

We use the MATCH software package \citep{dolphin_numerical_2002, dolphin_estimation_2012, dolphin_estimation_2013} to characterize the stellar MF for a young star cluster. For a given evolution library, MATCH constructs a simple stellar population which is then convolved with completeness, noise, and photometric biases derived from artificial stars tests (ASTs). The resulting synthetic CMD is then linearly combined with a suitable CMD of foreground/background populations and scaled to best reproduce the observed cluster CMD. The fit quality is calculated according to a Poisson likelihood, fit over a large grid of parameters, including IMF, to fully cover all likelihood space. 

For this analysis, we adopt Padova stellar evolution models \citep{girardi_acs_2010, marigo_evolution_2008, marigo_new_2017}, and a binary fraction of 0.35 where the mass ratio of the stars is drawn from a uniform distribution. This assumed binary fraction is within reasonable uncertainties for consistency with studies on binary fraction in clusters \citep{elson_binary_1998, albrow_frequency_2022}. While recent studies suggest binary fraction increases with stellar mass \citep{offner_origin_2023}, we do not have the infrastructure available to adequately incorporate this dependence into our study. Additionally, changes in the binary fraction have shown to have little effect on IMF slopes measured with the CMD modeling used here \citepalias{weisz_high-mass_2015}. We experimented with binary fractions ranging from 0.2 to 0.6, and our results were not significantly different. 

We adopt a distance modulus of 24.67 \citep{de_grijs_clustering_2014}, and assume a metallicity [M/H] of $-0.05 \pm 0.15$ consistent with the range of present-day gas phase metallicity in M33 \citep[e.g.,][]{rogers_chaos_2022}. To model the non-cluster background stars, we adopt the practice of \citet{johnson_panchromatic_2022} and construct an empirical CMD from an annulus around the cluster extending from $1.2-3.4$ times the photometric aperture. These sky annuli robustly characterize the sky flux levels and associated uncertainties with stellar contaminants. 

We model the cluster parameters as follows: MF slope ($\Gamma$) for stars $\geq 1$ $M_{\odot}$ from $-$1.2 to 5.0 with a step-size of 0.2 dex, log(Age/yr) from 6.6 to 9.0 with a step-size of 0.1 dex, extinction ($A_V$) from 0.0 to 2.5 mag with a step-size of 0.05 mag, and metallicity from $-$0.2 to 0.1 with a step-size of 0.1 dex. Resulting model CMDs are then compared to observed CMDs to determine the likelihood \citep{dolphin_numerical_2002}. 

\begin{figure*}[ht]
    \centering
    \includegraphics[width=0.90\textwidth]{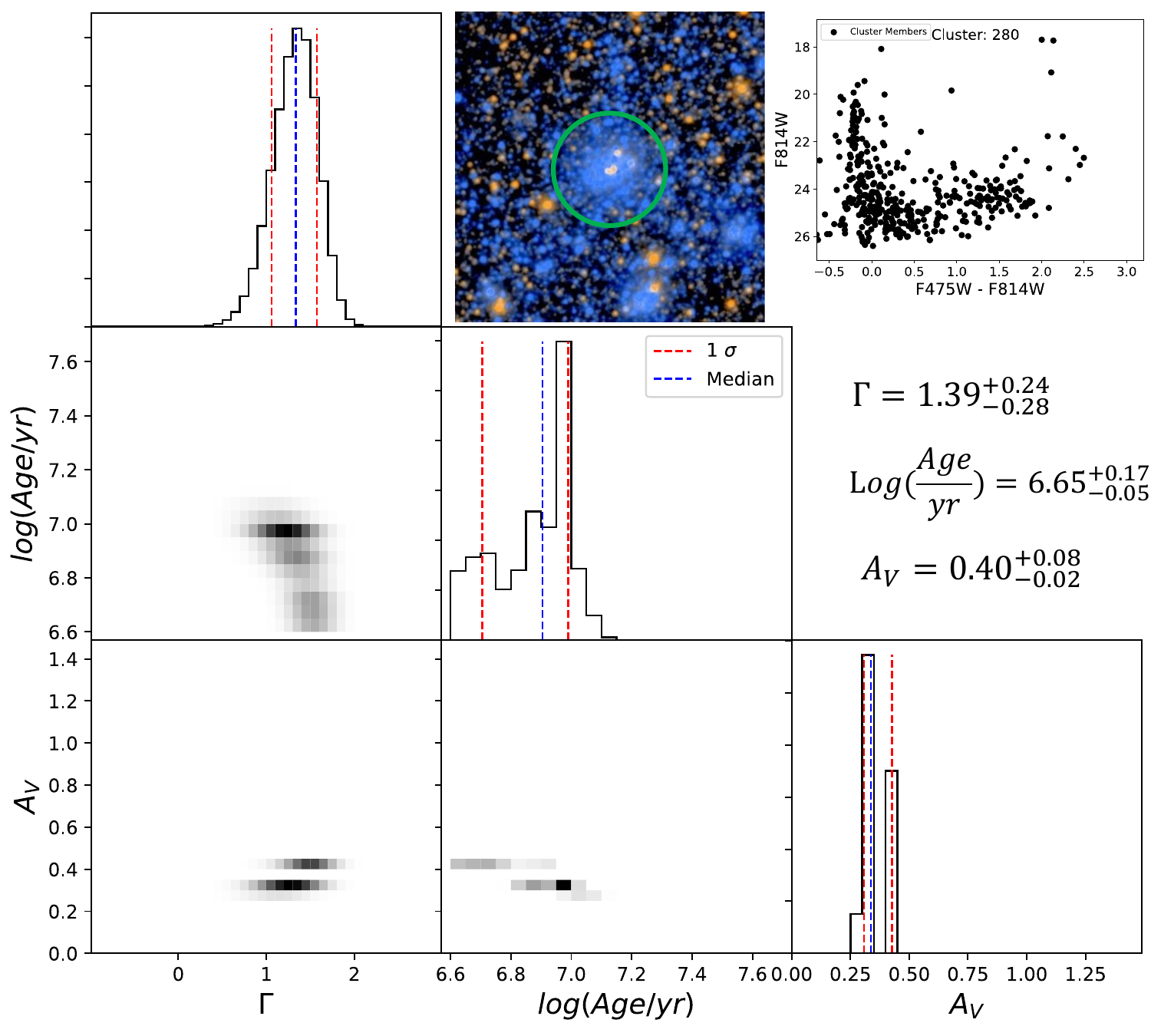}
    \caption{The marginalized PDF for M33 Cluster 280 showing the the posterior PDFs for $\Gamma$, age, and extinction. We also include the F475W and F814W combined color image with the 2.33 arcsecond radius visualized by the green circle. We also show the CMD, and best-fit age and $A_V$ estimates, along with the uncertainties, from \citet{wainer_panchromatic_2022} in the text which are interesting to compare to our full PDFs. We also show in the text the measured $\Gamma$. In the upper left, middle, and lower right panels, the blue line represents the median of the marginalized PDF, and the red lines correspond to the 16th and 84th percentiles. The bin sizes in the PDFs correspond to the grid with which each parameter was fit over, as described in the text.}
    \label{fig:cluster_pdf}
\end{figure*}

In Figure~\ref{fig:cluster_pdf}, we plot the joint and marginalized PDFs for three cluster parameters, $\Gamma$, age, and $A_V$. While this methodology does not account for every source of systematic uncertainty the parameters are tightly constrained, and the posterior probabilities give a reliable estimate of the uncertainty in the measured IMF \citep{weisz_panchromatic_2013}. 

Figure~\ref{fig:cluster_pdf} also shows that the IMF slope and the cluster age have correlated and non-Gaussian PDFs. We also see that the $A_V$ distribution is bimodal, which results from the combination of the common degeneracy between age and extinction and from the discreteness of the color distribution of stars in the cluster CMD. However, because we marginalize over age and $A_V$, the uncertainty driven by such correlations are incorporated into the total uncertainty on the IMF slope (upper left plot). This propagation of uncertainty is a particular strength of our technique, which accounts for degeneracies among several cluster properties.

\begin{figure}
    \centering
    \includegraphics[width=0.45\textwidth]{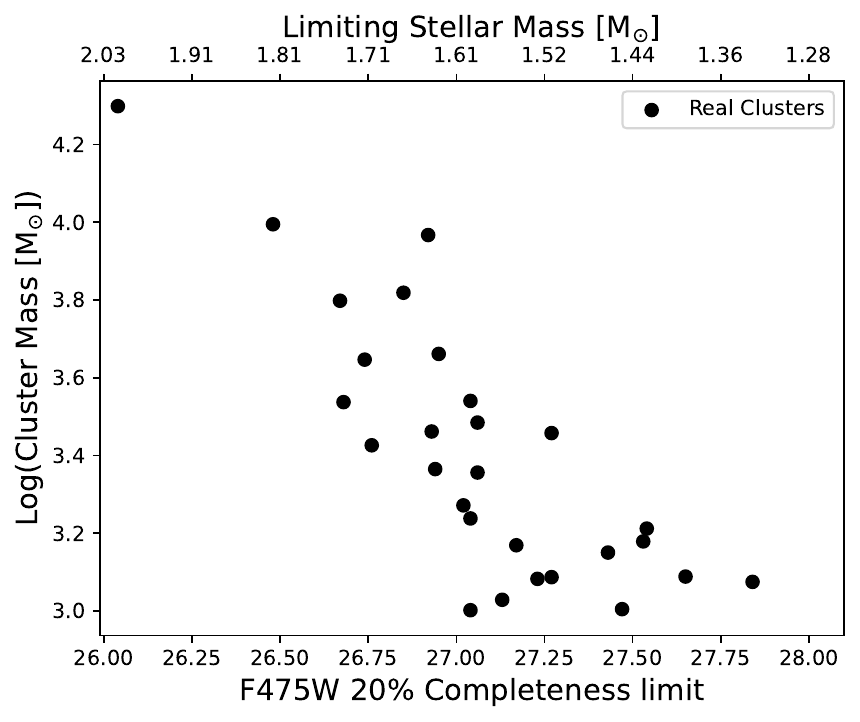}
    \caption{The $20\%$ F475W photometric completeness limit, as determined by ASTs, as a function of stellar mass. The secondary axis corresponds to the limiting main sequence stellar mass probed at this magnitude.}
    \label{fig:completeness}
\end{figure}

The fitted CMDs are comprised of all stars inside the \citet{johnson_panchromatic_2022} photometric aperture, which are modeled as a combination of cluster stars and background field stars. The magnitude range of the fitted CMD is determined based on cluster specific photometric completeness results derived from ASTs. We tested a wide range of completeness limits, from $5\%$ to $80\%$. From $20\%$ to $60\%$ completeness, our resulting $\Gamma$ remains unchanged. We find below $20\%$ completeness, the artifacts present in the photometry have a significant impact on the CMD and resulting IMF, while above $60\%$ there are too few stars to accurately constrain the IMF. We adopt the $20\%$ completeness limit which for synthetic clusters, was able to the best recover the input mass of the clusters. These completeness limits are presented in Figure~\ref{fig:completeness}, along with the corresponding main sequence stellar mass limit. From Figure~\ref{fig:completeness}, we can see a clear trend between photometric completeness and cluster mass. This trend in unsurprising as the more massive clusters have more stellar crowding, making resolving individual stars difficult. Note that the equivalent main sequence mass completeness limits for all clusters are $>$1 M$_{\odot}$, which is consistent with the assumption of a single high-mass power-law slope over the fitted mass range, whereas any turnover in the IMF \citep[e.g.,][]{kroupa_variation_2001} occurs at smaller stellar mass. 

\subsection{Inferring the IMF Slope from the Distribution of CMFs}
\label{sec:Bayesian_Step_2}

In this section we describe how we model the mean slope of the IMF, $\overline{\Gamma}$, and the intrinsic scatter in that slope, $\sigma_{\Gamma}$ from the ensemble of cluster MFs, obtained in Section~\ref{sec:Step1_MF_slope}. This step is to infer population-wide characteristics of the distribution of MF slopes, $\overline{\Gamma}$. To fit a Gaussian mixture model likelihood function for the MF of the clusters, we follow the approach laid out in \citet{hogg_data_2010} and \citet{foreman-mackey_exoplanet_2014}, using the affine invariant ensemble Markov chain Monte Carlo (MCMC) program \textsc{emcee} \citep{foreman-mackey_emcee_2013}. We closely follow the implementation of this methodology by \citetalias{weisz_high-mass_2015}, and a brief explanation follows.

The Gaussian mixture model framework allows us to consider that each data point has an associated probability of actually being a cluster, $Q_i$. We can then write down a Gaussian mixture model likelihood function for the MF of $i$ clusters, mitigating the impacts of questionable cluster candidates\footnote{We note that the exclusion of this parameter, giving each cluster equal weight, has very little impact on the resulting $\Gamma$ because of the high statistical probability that all of the clusters in our sample are truly clusters. We choose to include it to best replicate \citetalias{weisz_high-mass_2015}.}. Specifically, the mixture model can be written as 
\begin{equation}\label{eq:mf}
    p(\Gamma_i | \theta, Q_i) = Q_i \exp \left( -\frac{(\Gamma - \Gamma_i)^2}{2\sigma_{\Gamma}^2} -0.5 \log(2\pi \sigma_{\Gamma}^2) \right),
\end{equation}
where $\Gamma_i$ is the marginalized PDF for the MF of the $i$th cluster, and $\Gamma$ and $\sigma_{\Gamma}^2$ represent the mean and variance of the Gaussian that characterizes the high-mass slope. $\overline{\Gamma}$ is the mean slope for the true distribution, and $\sigma_{\Gamma}$ characterizes the intrinsic spread, such that small values of $\sigma_{\Gamma}$ indicate the IMF is more consistent with a universal IMF, with the sample variance being driven by observed uncertainty. This model allows us to not make any assumptions about the structure of the PDFs, considering the likelihood at all values, and incorporating the degeneracies from all grid runs. 

From this model, we can test the dependence of the MF slope on cluster properties such as age, mass, and radius. In order to test potential environmental dependencies on the IMF, we add two additional properties compared to \citetalias{weisz_high-mass_2015}, the local SFR in the region of the cluster, and the clusters galactic radius. The generalized model takes the form
\begin{equation}\label{eq:dependencies}
    \begin{array}{l}
        \Gamma(T_i, M_i, R_i) = \overline{\Gamma} + a_m \log\left(\frac{M_i}{M_c}\right) +a_t \log\left(\frac{T_i}{T_c}\right) \\+ a_r \log\left(\frac{R_i}{R_c}\right) + a_{sfr} \log\left(\frac{SFR_i}{SFR_c}\right) + a_{gr} \log\left(\frac{Gr_i}{Gr_c}\right)
    \end{array}    
\end{equation}
where $T_i$, $M_i$, and $R_i$ are the most likely age mass and radius of the $i$th cluster according to the \citet{wainer_panchromatic_2022} fits. $SFR_i$ and $Gr_i$ are calculated for each cluster using the \citet{lazzarini_panchromatic_2022} SFR maps, and the angular galactic radius assuming a common distance modulus. $T_c$, $M_c$, $R_c$, $SFR_c$ and $Gr_c$ are the mean of the full cluster sample in each case. Each parameter ($\overline{\Gamma}, \sigma_{\Gamma}^2, a_t, a_m, a_r, a_sfr, a_gr$) are assumed to have flat priors over the ranges reported in Section~\ref{sec:Step1_MF_slope}, where $\sigma_{\Gamma}^2$ must be greater than zero, and $a_t, a_m, a_r, a_sfr, a_gr$ must be less than one. These parameters are then sampled with the affine invariant ensemble MCMC program \textsc{emcee} \citep{foreman-mackey_emcee_2013}. 

For our MCMC calculation, we use 100 walkers, each performing 5000 steps, of which we discard the first 200 burn-in steps. We assess convergence of our chains according to the autocorrelation time, estimated to be $\sim60$ steps, which is exceeded by the burn-in period. In Section~\ref{sec: results} we report the median value of the marginalized posterior probability function for each of the seven parameters in the mixture model. 

\subsection{Reliability of Techniques, and Testing the Robustness of Results}\label{sec: testing_techniques}

In addition to running our full analysis on a set of synthetic clusters, we test critical steps of our technique through a quality test.  

\begin{figure*}[ht!]
    \centering
    \includegraphics[width=0.97\textwidth]{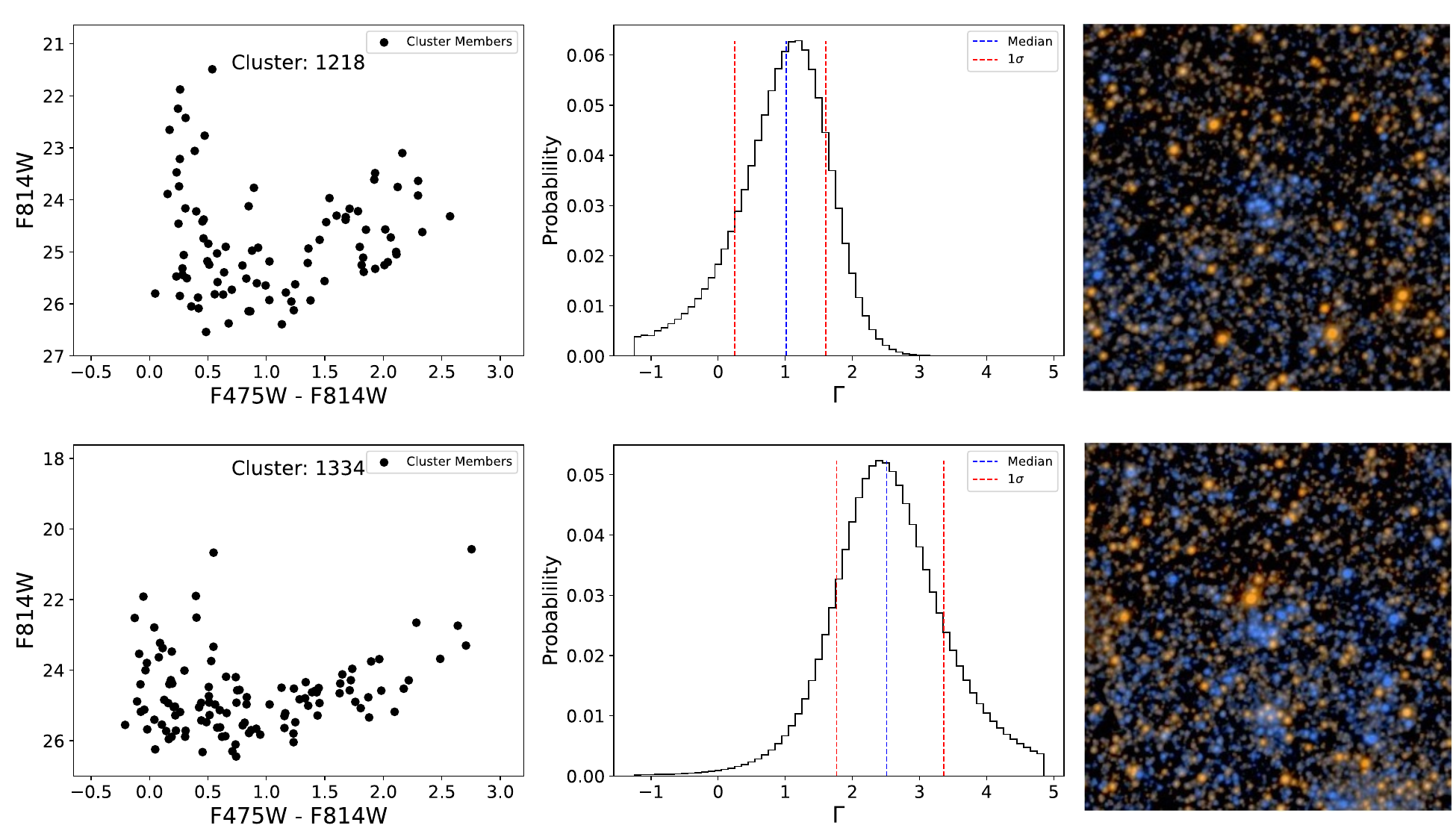}
    \caption{Shown are cluster CMDs, the $\Gamma$ marginalized PDFs, and the F745W and F814W color images for two low-mass clusters: clusters 1218 and 1334. The left panels are CMDs as reported in \citet{johnson_panchromatic_2022}, where the black points are cluster members. The middle panels are the marginalized $\Gamma$ PDFs for each cluster, where the blue line is the median and the red lines represent the 16th and 84th percentiles. }
    \label{fig:comp_pdfs}
\end{figure*}

First, we generate distributions of fake cluster PDF's and test the Gaussian Mixture methodology from Section~\ref{sec:Bayesian_Step_2}. We generate PDF's with medians randomly drawns between 0.35, and 2.35, with $1\sigma$ widths between 0.1 and 1.0 dex to represent with the typical observed uncertainty in $\Gamma$, and center the distribution of PDF's at $\Gamma$ values of 0.5, 1.3, and 2.0. For each case, we are able to recover the input value of the distribution within the uncertainties, demonstrating the mixture model technique is robust to different $\Gamma$ values.

As mentioned above, \texttt{MATCH} incorporates a background CMD into the fitting, and therefore should be reliable against single star outliers. To test this assumption, we perform a second experiment to quantify the impact a single massive star has on our results. Because the impact of single star outliers will be  most pronounced in low mass clusters, we select two low mass clusters, clusters 1218 and 1334, with $\log(M/M_{\odot})$ of 3.08 and 3.00, respectively. Both cluster 1218 and 1334 are of similar ages with $\log(\mathrm{Age/yr}) = 7.35$ and  $7.15$ \citep{wainer_panchromatic_2022}. However, the range in measured median MF slopes for these clusters spans more than 1.5 dex. These clusters are shown in Figure~\ref{fig:comp_pdfs}, where the left panel shows the CMDs, and the middle panel is the marginalized $\Gamma$ PDF, along with the color image of each cluster. 

For these low mass clusters, we perform an experiment testing this assumption, adding and subtracting a single bright star from the observed CMD, and investigate the results. For each cluster, we remove the brightest star in F814W. To add a bright star, we give the new star a F475W-F814W color equal the the brightest star in the cluster, and make the star 0.5 magnitudes brighter in F814W than the brightest star in the cluster. We find neither adding nor removing a star has an impact of $\Delta\Gamma > 0.04$ on the median measured MF. These results suggest that our measurements are not driven by the presence or absence of a single high mass star, but rather encompass the uncertainty in CMD scatter, and potential cluster contaminants. 

We also note the degeneracy between the fit cluster age and $\Gamma$, as seen in Figure~\ref{fig:cluster_pdf}. At the lowest cluster masses, there are typically larger age uncertainties, and a single high mass star can have an impact on the fit age, and thus effecting the $\Gamma$ likelihoods. However, this experiment indicates that even in the presence of this degeneracy, the age uncertainty is typically well incorporated in the $\Gamma$ PDF, and a single star won't have a statistically significant effect of the resulting $\Gamma$ measurement.

\subsection{Synthetic Clusters}
\label{sec:synthetics}
To test the end-to-end reliability of our analysis, we repeated our entire fitting technique (Section~\ref{sec: analysis}) for a set of synthetic clusters with known ages, masses, radii, and IMF slopes. Using \texttt{MATCH}, we generated 76 synthetic clusters spanning a range across the age, mass, and radii ranges of the real clusters. Individual stars for the synthetic clusters were drawn from a \citet{kroupa_variation_2001} IMF with a 0.35 binary fraction, with stellar properties using Padova stellar evolution models \citep{girardi_acs_2010, marigo_evolution_2008, marigo_new_2017}, and spatial positions drawn from a \citet{king_structure_1962} profile. 

We place the 76 synthetic clusters selected for analysis in a PHATTER field (B02-F08) where the number density of main sequence stars equals the median found for the sample of real clusters described in Section~\ref{sec: data} ($N_{\mathrm{MS}} \sim 2000$ for surrounding region; see Section 2.3 from \citealt{wainer_panchromatic_2022}). This is notably higher than the average for the PHATTER footprint ($N_{\mathrm{MS}} \sim 700$), which is not surprising as the youngest clusters are likely to be located in more dense star-forming regions. The insertion was performed by DOLPHOT into the optical F475W and F814W exposures that were used for cluster photometry, and run through the same photometry routines as the original data. The synthetic clusters were positioned by hand to avoid existing clusters, chip gaps, foreground stars, and background galaxies. The photometry catalogs were then processed the same way as for the real data, with cluster photometry being extracted from within an user-defined photometric aperture sized to mimic that of the real clusters. The surrounding field population is sampled using an annulus that spans $1.2-3.4$ times the photometric radius, 10 times the area of the cluster. Upon visual inspection of the inserted synthetic clusters, we removed 17 clusters that were not visually distinguishable from the field as they would not be in our real cluster sample which was determined from a visual search.

These synthetic clusters facilitated a variety of tests of our analysis routines. First, we take the populated isochrones of the synthetic clusters, and before inserting them into the images, we run our fitting technique on the isochrones and recover $\overline{\Gamma} = 1.30^{+0.02}_{-0.02}$, which is consistent with the input Kroupa IMF. This test demonstrates our ability to accurately measure the MF without observational noise, and the reliability of our methods described in Section~\ref{sec:Bayesian_Step_2}. It is also important to note that not every isochrone had a median MF slope equal to the input Kroupa IMF (16th to 84th percentile spread of 0.97 dex\footnote{The spread in median cluster $\Gamma$ is larger for lower mass clusters: $\sim1.3$ for clusters smaller than log(Mass/ M$_{\odot}$) of 3.6), and $\sim0.6$, for those larger, demonstrating more uncertainty at smaller masses.}), and only from looking at the ensemble of MF slopes were we able to fully recover the input. This result demonstrates that even in a regime without measurement uncertainty, Poisson statistics make drawing conclusion from single cluster measurement difficult. 

\begin{figure}
    \centering
    \includegraphics[width=.49\textwidth]{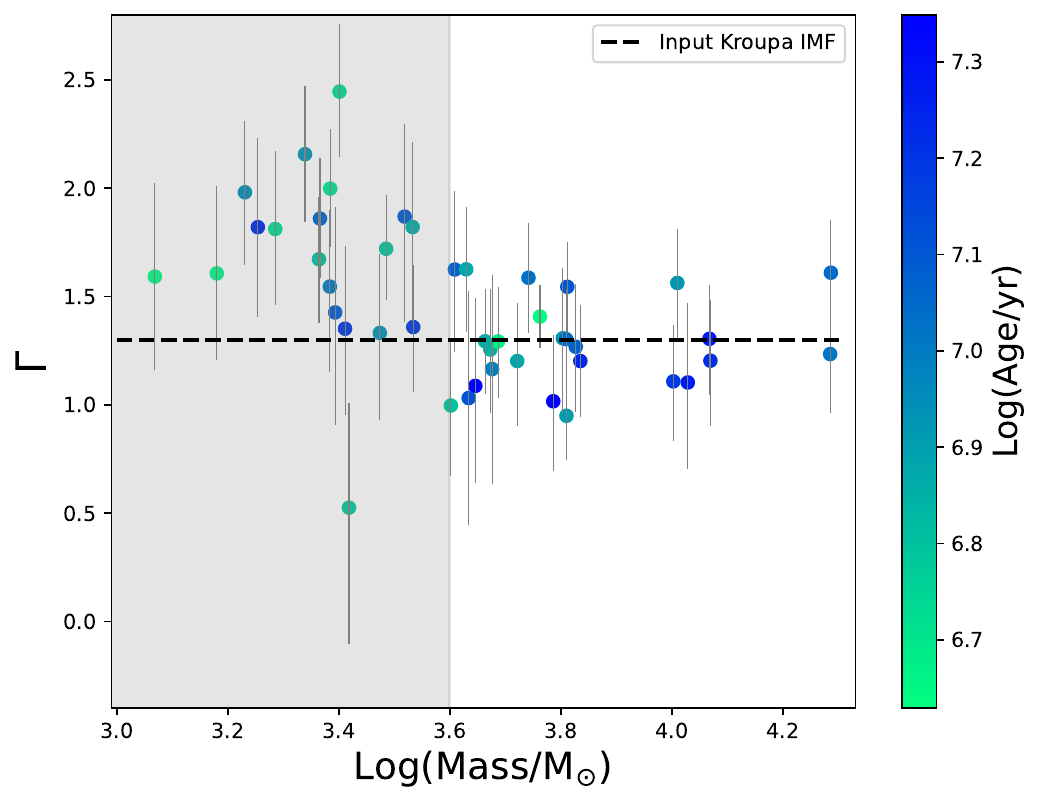}
    \caption{MF recovery for 45 synthetic clusters using the analysis techniques described in Section~\ref{sec: analysis}. The x-axis is the input mass of the synthetic cluster, and each point is colored by the input log(Age/yr) of the synthetic cluster. The black line represents the input Kroupa IMF slope. We can see a clear gradient in recovered $\Gamma$ with Mass, lower mass clusters recovering higher $\Gamma$ values. The grey region in the plot signifies the break at log(Mass/ M$_{\odot}$) of 3.6, where the synthetics recover the input IMF. }
    \label{fig:syn_gamma_vs_mass}
\end{figure}

From this point forward, the synthetic clusters are analyzed with the same software and in the same way as the real clusters. The individual MF results for the 45 synthetic clusters that passed the selection criteria from Section~\ref{sec: data} are shown in Figure~\ref{fig:syn_gamma_vs_mass}. It is clear from the figure that our analysis routine introduces a dependence on the recovered MF slope with the input mass of the cluster. When fitting the ensemble for the full swath of synthetics, we verify this dependence by recovering a $3\sigma$ $a_m$ of $-0.21\pm{0.07}$.
The figure shows a clear systematic bias toward measuring higher $\Gamma$ at low masses. 

We investigate different mass regimes where we can reliably recover the input Kroupa IMF. From Figure~\ref{fig:syn_gamma_vs_mass}, we divide the synthetic sample into two sub-samples, above and below log(Mass/ M$_{\odot}$) of 3.6 (26 and 19 clusters, respectively), and find the lower mass sample shows a significant offset to a higher slope value, with an ensemble value of $\overline{\Gamma} = 1.76^{+0.10}_{-0.11}$.

\begin{figure}
    \centering
    \includegraphics[width=0.48\textwidth]{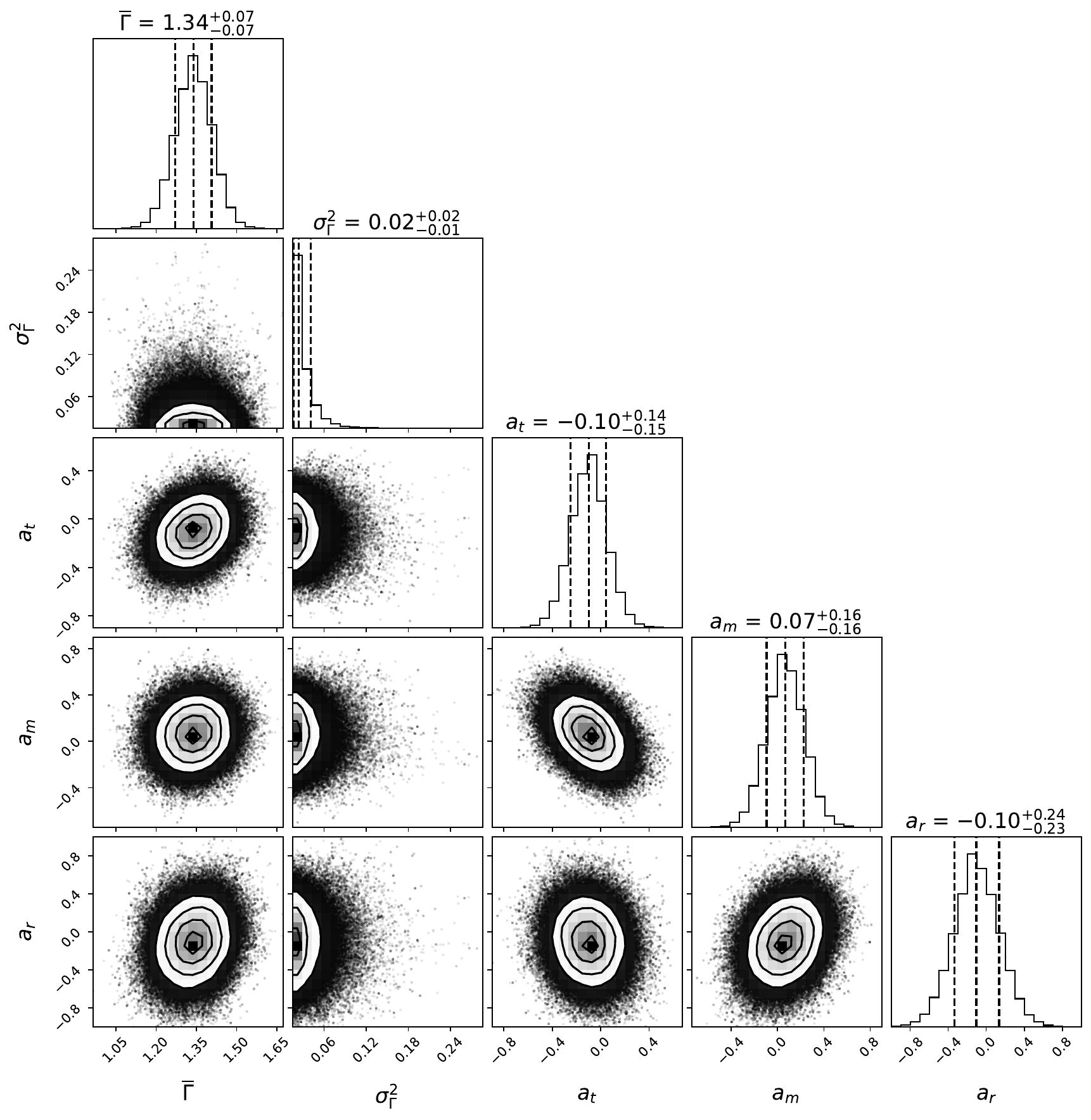}
    \caption{The joint and marginalized distributions for the distribution of synthetic cluster MF for 26 synthetic clusters with masses larger than 3.6 log(Mass/M$_{\odot}$. In all panels, the dashed lines represent the 16th, 50th, and 84th percentiles of each marginalized distribution. }
    \label{fig:syn_corner}
\end{figure}

We have extensively tested this bias to high $\Gamma$ for low-mass clusters: performing quality tests on our ASTs, distributing ASTs according to an empirical effective radius opposed to the half light radius, adjusting photometric radii, fitting over different magnitude ranges, using a fixed the limiting magnitude for the fitted CMD, and inserting the synthetic clusters in less dense regions. However, the bias at low masses persists. We conclude that at lower masses, the scatter in MF slopes is more drastic, and the synthetic cluster results suggest that fitted $\Gamma$ values have the potential to be biased high. For low mass clusters, the increased scatter in fitted MF slopes is likely due to the decreasing ratio of cluster stars to field stars, leading to confusion in the faint regions of the CMD and degeneracies in the fitting. In M33, the field is significantly more dense than in M31, with the average number of main sequence stars in PHATTER images a factor of 5 greater than in PHAT \citep{johnson_panchromatic_2022}. For low mass clusters, where star counts are smallest, the background component becomes more significant, and may therefore be at least partially to blame for the bias seen in the synthetic clusters results which was not seen in \citetalias{weisz_high-mass_2015}. 

However, for clusters more massive than log(Mass/ M$_{\odot}$) of 3.6, we find our methods accurately recover the input IMF. In the synthetic sample, there are 26 clusters above the mass limit. The results for these synthetic 26 clusters in this range are shown in Figure~\ref{fig:syn_corner}. This higher mass sample recovers a mean IMF slope of $\overline{\Gamma} = 1.34\pm0.07$ that is consistent with the input IMF and shows no significant dependencies, as presented in Figure~\ref{fig:syn_corner}. Therefore, we establish a "gold" sample, which represents clusters above log(Mass/ M$_{\odot}$) of 3.6, where we expect little to no mass dependence in our results. These results demonstrate that for log(Mass/ M$_{\odot}$) above 3.6, we are able to corroborate our results against synthetic clusters, and verify the reliability of our analysis techniques.

In the M33 cluster sample (Section~\ref{sec: data}), we find there are 9 clusters more massive than the synthetic-derived ``gold'' reliability threshold of log(Mass/ M$_{\odot}$) of 3.6. To better quantify the effect that small number statistics of this ``gold'' sample have on the robustness of our results, we perform a final test on the synthetic sample. In order to test if there is an implicit bias from selecting a subsample of clusters from a parent distribution, we randomly selected 9 of the 26 synthetics above log(Mass/ M$_{\odot}$) of 3.6 to match the number of real clusters we observe above. We perform 1000 draws without replacement in each trial selecting a sub-sample of 9 synthetic clusters. For the 1000 trials, we find the median bootstrapped $\overline{\Gamma} = 1.34\pm0.07$, with each trial having an ensemble uncertainty of $0.27\pm0.02$, around $4\times$ larger than the uncertainty for the sample of 26 clusters, which is consistent with our expectations based on the scaling of uncertainty with N$_{clusters}$ from \citetalias{weisz_high-mass_2015}. Moreover, the dependence parameters $a_t$, $a_m$, and $a_r$, have median values of $0.00\pm0.18, 0.00\pm0.017,$ and $0.00\pm0.022$ respectively for the 1000 trials. 

These results are consistent with the ensemble values we recovered for the larger sample of 26 clusters shown in Figure~\ref{fig:syn_corner}, demonstrating that having a sub-sample of only 9 clusters does not introduce additional biases from the ensemble values of a larger sample. These results also validate the $\sigma_{\Gamma}$ value reported in Figure~\ref{fig:syn_corner}, indicating that over this gold sample of synthetics there is little intrinsic spread in MF values, consistent with a universal IMF. This is reassuring, as all synthetic clusters were generated with the same IMF, and establishes a floor for our ability to detect a truly universal IMF in the real data. 

\section{Results}\label{sec: results}

In this section, we present the results for the IMF in M33, described by the two steps in Section~\ref{sec: analysis}. First, we present the present-day MF slopes for our full sample of 34 clusters, and then discuss the results for the 9 clusters in the gold sample range described in Section~\ref{sec:synthetics}.

\begin{figure*}[!ht]
    \centering
    \includegraphics[width=.99\textwidth]{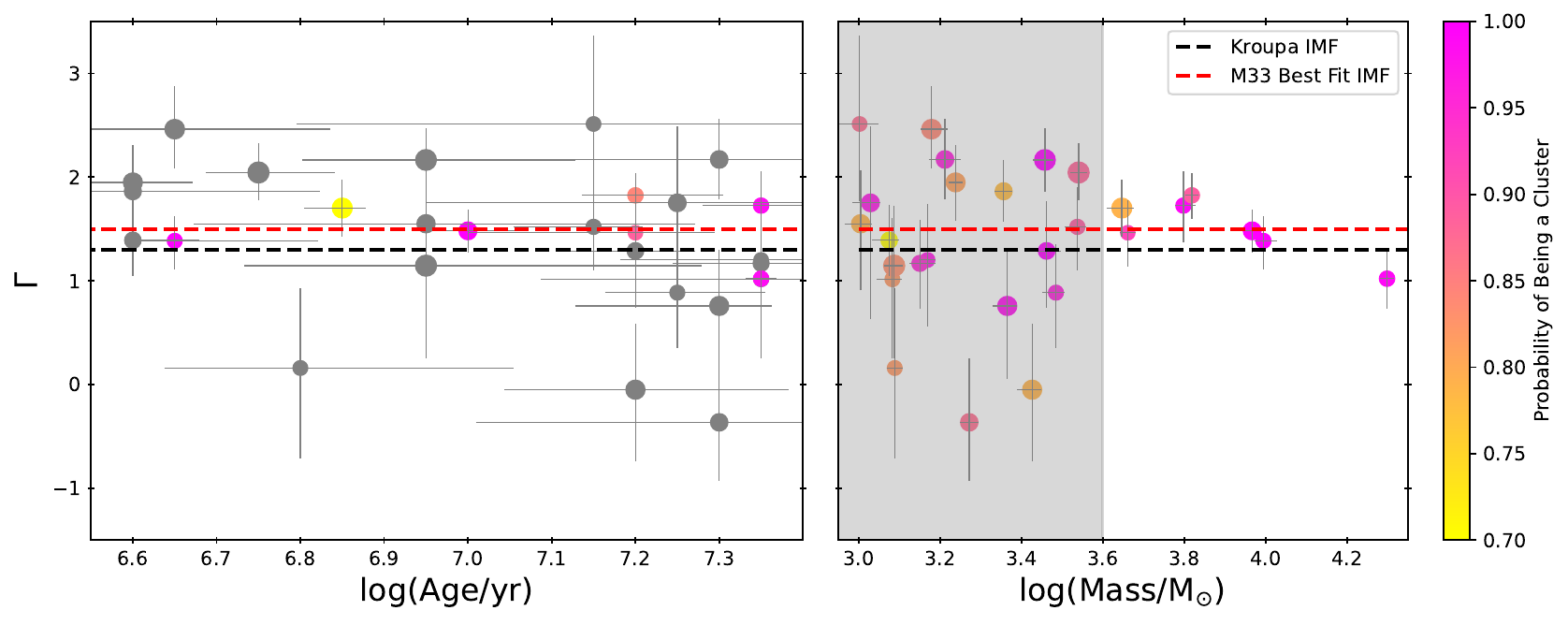}
    \caption{The MF slope for each cluster compared to the best fit cluster age (left) and mass (right) from the \citet{wainer_panchromatic_2022} fits. Each point size is scaled by the cluster radius and colored by the probability of being a cluster based on the weighting system of \citet{johnson_panchromatic_2022}. The black dashed line is at the Kroupa IMF of $\Gamma = 1.3$, and the red line shows the median IMF result from this work of $\Gamma = 1.49$. The greyed region represents the region of mass space where our synthetic clusters suggest a potential for bias. The left panel has points that are lower than 3.6 log(Mass/M$_\odot$) greyed out.}
    \label{fig:clstMF_vs_age_mass}
\end{figure*}

The individual MF results for our full sample of clusters are shown in Figure~\ref{fig:clstMF_vs_age_mass}. In this figure, we compare the fitted MF slope versus the cluster age and mass as presented in \citet{wainer_panchromatic_2022}. The black line in the figure represents the \citeauthor{kroupa_variation_2001} IMF slope of $\Gamma = 1.30$. The greyed out region in the right panel represents the range of masses where our synthetic clusters indicate a potential for bias. Below this limit of log(Mass/ M$_{\odot}$) of 3.6, we can visually see the spread in MF values increase; however there does not appear to be a systematic bias to higher values as we saw for the synthetic clusters.
While the reason for the lack of mass-dependent behavior in the observed sample is unclear, it suggests there could be fundamental differences between real and synthetic clusters, such as unrealistic synthetic photometry of cluster members or the cluster's spatial profile. All 34 cluster MFs are presented in Table~\ref{tab:clst_gammas}, along with all the properties which go into Equation~\ref{eq:dependencies}.

\begin{deluxetable*}{cccccccccc}
\tabletypesize{\small}
\setlength{\tabcolsep}{0.05in}
\tablewidth{0pt}
\tablecaption{Individual Cluster MF Results \label{tab:clst_gammas}}

\tablehead{
\multirow{2}{*}{ID} & \multirow{2}{*}{Gamma 16} & \multirow{2}{*}{Gamma 50} & \multirow{2}{*}{Gamma 84} & \colhead{Age\tablenotemark{a}} & \colhead{Mass\tablenotemark{a}} & \colhead{AV\tablenotemark{a}} & \colhead{Clst Radius\tablenotemark{b}}, & \colhead{SFR\tablenotemark{c}} & \colhead{Gal Radius}\\ & & & & [log(Age/yr)] & [log(Mass/M$_{\odot}$)] & [mag] & [parsec] & [log(M$_{\odot}$ yr$^{-1}$ kpc$^{-2}$)] & [kpc] }

\startdata
45 & 1.22 & 1.43 & 1.64 & 7.0 & 3.97 & 0.35 & 2.73 & -3.61 & 0.7 \\
163 & 2.03 & 2.41 & 2.83 & 6.65 & 3.18 & 1.05 & 3.32 & -3.43 & 2.47 \\
179 & 1.32 & 1.67 & 2.0 & 7.35 & 3.8 & 0.4 & 1.97 & -3.5 & 1.09 \\
200 & 1.08 & 1.41 & 1.72 & 7.2 & 3.66 & 0.45 & 1.76 & -3.98 & 1.32 \\
280 & 1.06 & 1.34 & 1.58 & 6.65 & 3.99 & 0.4 & 1.89 & -3.15 & 1.69 \\
\enddata
\tablecomments{Table \ref{tab:clst_gammas} is published in its entirety in the electronic edition of the {\it Astrophysical Journal}. A portion is shown here for guidance regarding its form and content.}
\tablenotetext{a}{Estimates in the table represent the 'Best-fit' parameters from \citet{wainer_panchromatic_2022}.}
\tablenotetext{b}{Estimates in the table represent the Half light radius ('$R_{eff}$') from \citet{johnson_panchromatic_2022}.}
\tablenotetext{c}{Estimates in the table are area normalized values of the SFR from 0-100 Myrs from \citet{lazzarini_panchromatic_2022}.}

\end{deluxetable*}

\begin{figure*}[ht]
    \centering
    \includegraphics[width=.95\textwidth]{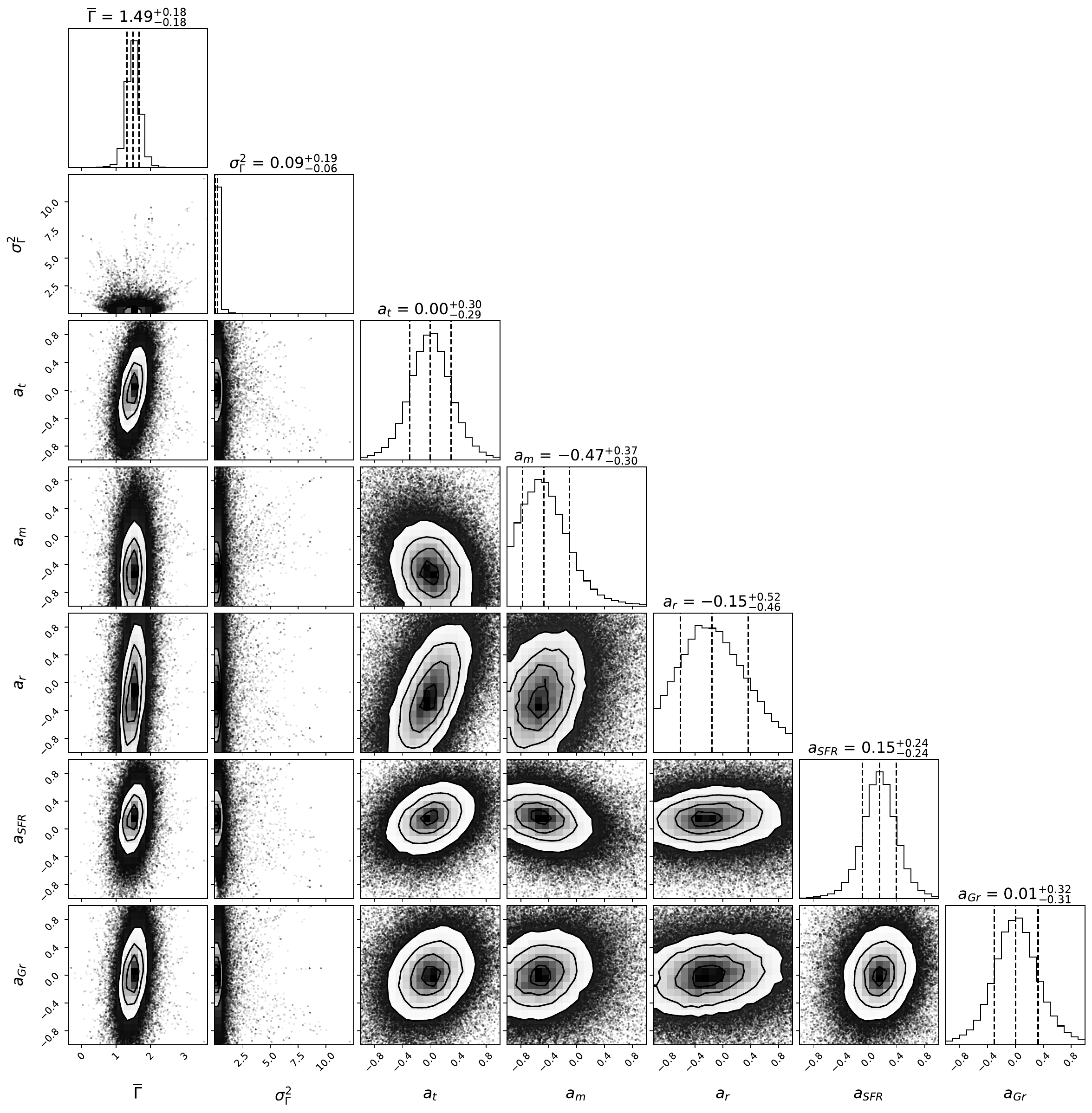}
    \caption{The joint and marginalized distributions for the distribution of cluster MF, described by Equations~\ref{eq:mf} and~\ref{eq:dependencies}. In all panels, the dashed lines represent the 16th, 50th, and 84th percentiles of each marginalized distribution. Parameters are the mean slope of the IMF, $\overline{\Gamma}$, and the intrinsic scatter in that slope, $\sigma_{\Gamma}$ from the ensemble of cluster MFs, and well and parameters testing for dependencies on cluster properties: cluster age $a_t$, mass $a_m$, cluster radius $a_r$, local SFR $a_{sfr}$, and galactocentric radius $a_{gr}$, as described in the text. }
    \label{fig:full_sample_corner}
\end{figure*}

In the second step of our analysis, we present the results for the Gaussian mixture model, reporting the results for the seven parameters of the model: the mean MF slope $\overline{\Gamma}$, the intrinsic dispersion $\sigma_{\Gamma}$, and the linear trend with the data against the cluster properties, cluster age $a_t$, mass $a_m$, cluster radius $a_r$, local SFR $a_{sfr}$, and galactocentric radius $a_{gr}$. These results are shown in Figure~\ref{fig:full_sample_corner}.  

As seen in Figure~\ref{fig:full_sample_corner}, the mean IMF slope is $\overline{\Gamma} = 1.49\pm0.18$. These results are consistent at the $1\sigma$ level with the \citeauthor{kroupa_variation_2001} IMF of $\Gamma = 1.3$ and the \citeauthor{salpeter_luminosity_1955} IMF of $\Gamma = 1.35$, but they are closer to the results from young clusters in M31, which were also steeper than these standard values \citetalias{weisz_high-mass_2015}. 

While we report the median and $1\sigma$ uncertainty of the $\sigma^{2}_{\Gamma}$ value in Figure~\ref{fig:full_sample_corner} for consistency with other parameters, the distribution is highly un-Gaussian, and limited by 0. Therefore, we follow the procedure of \citetalias{weisz_high-mass_2015}, and characterize $\sigma^{2}_{\Gamma}$ with the mode of the distribution, and quantify its uncertainty by the 68th percentile. With this nomenclature, we find the intrinsic scatter of the distribution $\sigma^{2}_{\Gamma}$ to be $0.02^{+0.16}_{-0.00}$. As demonstrated by \citetalias{weisz_high-mass_2015}, the ensemble uncertainty scales as $\propto 1 /N_{clusters}$. Because our gold cluster sample is small, with only 9 PHATTER clusters, a larger sample of clusters would provide a tighter constraint on the M33 IMF. 

Within the gold sample, we find no dependence of the IMF with cluster age, cluster radius, local SFR, or galactic radius, but do see a slight dependence with cluster mass at the $\sim 1\sigma$ level, where $a_m$ is $-0.47^{+0.37}_{-0.30}$. However, this mass dependence can easily be explained by the very limited number of clusters in the gold sample. Primarily, this correlation is driven by the single cluster at log(Mass/ M$_{\odot}$) $= 4.3$. Given the small number statistics, and without any additional data above log(Mass/ M$_{\odot}$) $= 4.0$, any correlation depends heavily on this single cluster's $\Gamma$ measurement. Therefore, while the gold sample has a small dependence on cluster mass, we believe it is likely not a physical effect, but rather a result of small number statistics, and more clusters are required to make any significant claims.

In addition to the gold sample, we also calculate the ensemble IMF for the full sample of 34 clusters. For the full sample, we find $\overline{\Gamma} = 1.45\pm0.10$, consistent with the result for the gold sample. Importantly, over the full range of cluster masses, we find no mass dependence with $a_m = -0.01\pm 0.11$. We also note that in the left hand panel of Figure~\ref{fig:clstMF_vs_age_mass}, there appears to be a dependence on cluster age. In fact, when considering the the full sample of 34 clusters, there is a slight age dependence of $a_t = -0.27{\pm0.17}$. However, as seen in Figure~\ref{fig:full_sample_corner}, the gold sample of 9 clusters shows evidence of no age dependence with an $a_t = 0.00^{+0.30}_{-0.29}$. Moreover, when considering the slight age dependence in the full 34 clusters, it is important to note that it is driven by clusters which are less massive than the high-confidence threshold of log(Mass/ M$_{\odot}$) of 3.6. We iterate that more clusters are needed to make claims about any potential dependence.

While the mass-limited gold sample was conservatively chosen to avoid possible fitting biases revealed by synthetic cluster tests, we reiterate that $\Gamma$ results for the full observed sample show no evidence for cluster mass-dependent biases. We conclude there is no evidence that ensemble constraints on $\overline{\Gamma}$ depend on cluster sample selection, and that the full sample could benefit from improved precision due to larger number statistics, but adopt results for the gold sample as this work's primary result out of an abundance of caution.

\section{Discussion}\label{sec: discusion}
While our measurements of the IMF in M33 clusters is of interest for M33-specific studies, it is also of interest for constraining the dependence of the IMF on environment. Below, in Section~\ref{sec:clst_to_clst_var}, we discuss the cluster to cluster variance of the IMF, illustrating the need to perform IMF analysis on as large a sample of clusters as possible. We then explore environmental impacts on the IMF in Section \ref{sec:environment}, and discuss the broader implications a steeper IMF has on the number of high-mass stars created, and how that impacts the broader astronomical research enterprise in Section~\ref{sec:implications}.

\subsection{Cluster to Cluster Variance of the IMF }\label{sec:clst_to_clst_var}

In M33, for our gold sample, we report an ensemble scatter in MF slopes of $0.02^{+0.16}_{-0.00}$ (visualized in Figure~\ref{fig:full_sample_corner}), which is consistent with a universal IMF as determined by the synthetic clusters. However, when considering the more inclusive sample that extends down to log(Mass/M$_{\odot}$) of 3.0, we find there is more scatter, with an ensemble scatter of $0.04^{+0.13}_{-0.00}$, higher than what we expect from the synthetic cluster results. This scatter is also visible in Figure~\ref{fig:clstMF_vs_age_mass}, where in particular, the low mass clusters have a much higher spread in $\Gamma$ values. There is a well documented history of intrinsic scatter within the compilation of IMF slope measurements from individual clusters \citep{kroupa_initial_2002, lee_origin_2020}. We also see scatter from cluster to cluster in the IMF slope, which emphasizes the need for determining IMF properties from an ensemble of clusters, rather than single clusters. 

This variance is highlighted in our sample by the two clusters shown in Figure~\ref{fig:comp_pdfs}. Despite the two clusters being very close in age and mass, the number of high-mass main sequence stars is visibly different in the two CMDs, and the resulting $\Gamma$ values are therefore different. 

To explore the observed variation with respect to our uncertainties, we quantify the scatter in the distribution of MF slopes by determining the reduced $\chi^2$. For the full sample of 34 clusters, we find $\chi^2$ to be 1.98, while for the gold sample, $\chi^2$ is 1.33. This result demonstrates that the scatter in the sample at smaller masses is larger than the uncertainties would predict, which suggests intrinsic IMF scatter across the cluster sample. However, when we compare this distribution of $\chi^2$ values to a normal distribution through a Kolmogorov-Smirnov (KS) test, we recover a K-S statistic of 0.20 and a p-value of 0.21. Therefore, we are unable to conclusively claim that the variance between clusters is more than what would be expected from stochastic sampling of the IMF. This work highlights the need for large samples of clusters in order to make substantial claims about the IMF. We emphasize that drawing galactic, or universal, claims about the IMF from single clusters is extremely difficult. 

\subsection{Exploring Environmental Impacts on the IMF}\label{sec:environment}

As discussed through this paper, one of the most pressing questions in studies of the high-mass IMF is the universality of the IMF, or whether there are significant environmental effects which impact the number of high-mass stars formed \citep[e.g.,][]{lee_origin_2020}. Most notably, there has been recent evidence to support an IMF which is variable in extreme star forming environments or environments with different metallicities \citep{li_stellar_2023, rautio_constraining_2023, zhang_stellar_2018, marks_evidence_2012, weidner_top-heavy_2011}. We have discussed cluster to cluster variations of the IMF in our study (Section~\ref{sec:clst_to_clst_var}), and in this section, we will investigate potential impacts the local cluster environment, such as SFR and metallicity, has on the measured cluster MF, and inferred IMF. 

On a galactic scale, The $\overline{\Gamma} = 1.45^{+0.03}_{-0.06}$ measured in M31 by \citetalias{weisz_high-mass_2015} serves as the tightest systematically constrained measurement of a galactic high-mass IMF from resolved stars. The M33 ensemble IMF results of $\overline{\Gamma} = 1.49^\pm0.18$ of our gold sample, as well as the $\overline{\Gamma} = 1.46^{+0.11}_{-0.12}$ from the more inclusive sample, presented in Section~\ref{sec: results} are consistent with the M31 value, and adds an additional galactic data point using multiple clusters across galactic environments. This consistency between the derived $\Gamma$ values for M33 and M31 is of particular interest as M33 is a more active star forming galaxy, with an average star formation rate surface density log($\langle \Sigma_{SFR} \rangle$~/~(\Msun~yr$^{-1}$~kpc$^{-2})$) of $-2.04^{+0.16}_{-0.18}$ \citep{wainer_panchromatic_2022, lazzarini_panchromatic_2022}, nearly four times higher than M31 (log($\langle \Sigma_{SFR} \rangle$~/~(\Msun~yr$^{-1}$~kpc$^{-2})$) of $-2.68^{+0.26}_{-0.38}$ \citep{lewis_panchromatic_2015, johnson_panchromatic_2017}. Because of the higher $\Sigma_\mathrm{SFR}$, M33 is able to make slightly more massive clusters than M31 \citep{johnson_panchromatic_2017, wainer_panchromatic_2022}. Despite the difference in cluster masses and $\Sigma_\mathrm{SFR}$ between the two galaxies, the measured IMFs are the same. This consistency supports the universality of the mean high-mass IMF in the Local Group.

On an individual cluster scale, we investigate potential impact local cluster environments could have on a clusters MF. To quantify the SFR associated with each cluster in our sample, we use the SF maps of \citet{lazzarini_panchromatic_2022} who fit color magnitude diagrams to resolved stars in 2005, $\sim100$ parsec regions. Based on the cluster location, we adopt the \citet{lazzarini_panchromatic_2022} SFR value between 0-100 Myr ago as the cluster SFR. We normalize the value given in \citet{lazzarini_panchromatic_2022} by the area in order to estimate the $\Sigma_\mathrm{SFR}$, a value more comparable to galactic samples in the literature.

\begin{figure*}
    \centering
    \includegraphics[width=0.97\textwidth]{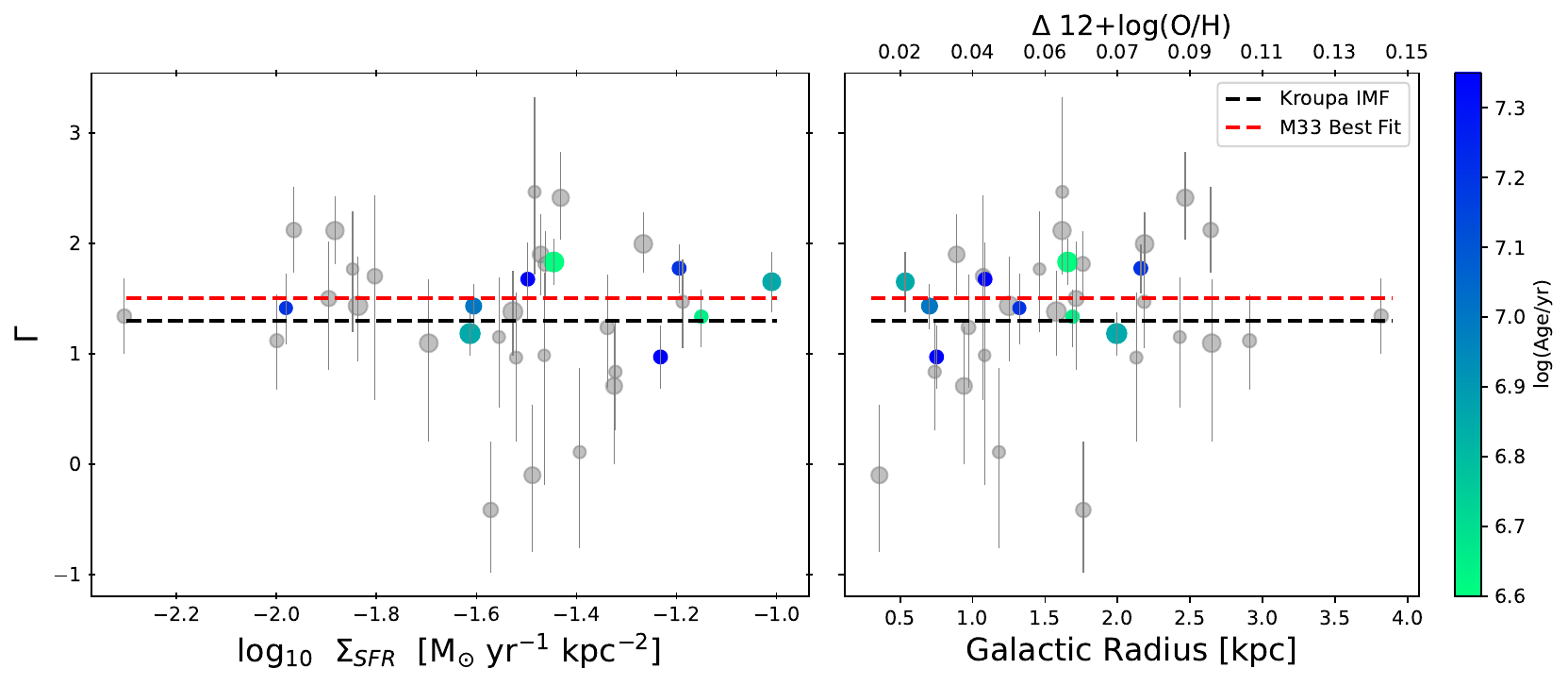}
    \caption{Cluster $\Gamma$ as a function of cluster $\Sigma_{SFR}$, and galactic radius, which serves as a proxy for metallicity. The $\Sigma_{SFR}$ is measured for a $\sim0.01$ kpc$^2$ region containing the cluster, calculated by \citet{lazzarini_panchromatic_2022} from 0-100 Myrs. Each point is colored by the log(Age/yr) of the cluster and is scaled by the radius of the cluster to easily map to Figure~\ref{fig:clstMF_vs_age_mass}, and greyed out if it is below log(Mass/M$_{\odot}$) of 3.6. These results show no trend in IMF with either $\Sigma_\mathrm{SFR}$, or galactic radius. The secondary axis for galactic radius is the change in 12+log(O/H) abundance gradient with radius according to \citet{rogers_chaos_2022}.}
    \label{fig:sfr_vs_gamma}
\end{figure*}

The individual cluster MF results are shown in the left panel of Figure~\ref{fig:sfr_vs_gamma}. As we can see, there is no significant trend with the cluster neighborhood SFR. Moreover, for the 9 clusters in our gold sample, visualized in Figure~\ref{fig:full_sample_corner}, we find no significant trends in our Gaussian mixture model with a $a_{SFR} = 0.15 \pm0.24$. These results show no dependence of the IMF on the SFR in the region around the cluster.  

It is no surprise that the majority of the newest formed star clusters are located in environments with higher than average $\Sigma_\mathrm{SFR}$. Therefore, while the increased $\Sigma_\mathrm{SFR}$ in these regions covers a wide range of the galactic $\Sigma_\mathrm{SFR}$ presented in \citet{leroy_phangs-alma_2021}, (from $-2.3 \leq \Sigma_\mathrm{SFR} \leq -1.0$) even the most extreme star forming regions in M33 are not representative of the extreme environments suggested to create a more top-heavy IMF \citep{zhang_stellar_2018}. While we do not see any IMF dependence on the $\Sigma_\mathrm{SFR}$ in our sample, due to its smaller mass, the galactic M33 $\Sigma_\mathrm{SFR}$ of $\sim-2$ \citep{wainer_panchromatic_2022, leroy_z_2019} is small compared to the majority of galaxies in the PHANGS survey \citep{leroy_phangs-alma_2021}. It would be of significant interest to the community to measure the IMF for a large sample of clusters with resolved stars in a star-bursting galaxy. 

The other primary environmental effect which has been discussed in the literature to affect the IMF is metallicity. While there is a gas phase metallicity gradient across the M33 disk, the PHATTER footprint is constrained to the five innermost kpc, and the metallicity does not begin to significantly change until the outer regions of the galaxy \citep{rogers_chaos_2022}. While within the PHATTER footprint the metallicity is fairly constant with a measured 12+log(O/H) abundance gradient of $-0.037 \pm 0.007$ dex kpc$^{-1}$ \citep{rogers_chaos_2022}, we nevertheless show the cluster $\Gamma$ results against galactic radius, which serves as a proxy for metallicity. These results are shown in the right hand panel of Figure~\ref{fig:sfr_vs_gamma}. 

From the measured 12+log(O/H) gas phase abundance gradient of $-0.037 \pm 0.007$ dex kpc$^{-1}$, the majority of our clusters are between 0.5 and 3 kpc, and therefore, across our sample of clusters, we expect a change of $\sim-0.1$ dex. Similarly, for the 9 clusters in our gold sample, we only see a difference of $\sim1.5$ kpc in Galactic radius, which is only a $\sim-0.06$ dex difference in metalicity. The results for the gold sample are shown in Figure~\ref{fig:full_sample_corner}, and we do not see a trend in our Gaussian mixture model with $a_{gr} = 0.01^{+0.32}_{-0.31}$, showing no significant dependence of the IMF on the galactic radius. It would be very interesting to have a survey covering the entirety of the M33 disk, and be able to have clusters in different metallicity environments. We also note that when measuring the environmental statistics for the full sample of clusters which extends to log(Mass/M$_{\odot}$) of 3.0, we do not see any dependencies, and values of $a_{sfr}$ and $a_{gr}$ are consistent with those of the gold sample.

Recent theoretical work has proposed additional tests of environmentally-dependent IMF behavior, however in practice these tests are not observationally feasible -- even with a high-quality dataset like PHATTER. For example, \citet{grudic_does_2023, grudic_great_2023} proposes a comparison between the maximum stellar mass found in a cluster and the total number of cluster stars. In practice, both the maximum mass and the total cluster mass are difficult to constrain due to observational magnitude and crowding limits. Moreover, both the number of stars in a cluster and the number of stars above 8 M$_{\odot}$ are age-dependent measurements. While we have preferentially selected the youngest clusters to probe the initial state of the cluster, it is very difficult to differentiate between an evolved system where the most massive stars have already evolved away and one where those most massive sources were never formed \citep[e.g.,][]{schneider_evolution_2015}. Observers have also used indirect tracers of massive stars to place constraints on the upper IMF \citep[e.g.,][]{jung_universal_2023}, however this path is also difficult.

In this study, we do not see any dependence on the IMF with $\Sigma_\mathrm{SFR}$ or 12+log(O/H). As noted above, the change in these two parameters across the PHATTER survey is relatively small. However, the outer regions of M33 see a significant extended HI disk that reaches far beyond the visible stellar disk \citep{koch_kinematics_2018}. Additionally, the metallicity gradient throughout the disk may affect the cooling efficiency of gas, which in turn influences the rate of star formation. Given the uncertainty scaling of the IMF with number of clusters in \citetalias{weisz_high-mass_2015}, if a high mass slope difference exists for different environments, of order $\sim40$ clusters is needed for a $3\sigma$ detection. Extended coverage of the M33 disk would provide substantial evidence for any potential impact SFR or metallicity has on the high mass IMF.

\subsection{Implications of a Steeper IMF}
\label{sec:implications}

The work presented in this paper is formally consistent with the canonical \citeauthor{kroupa_variation_2001} and \citeauthor{salpeter_luminosity_1955} values for the high mass slope of the IMF. However, it is also consistent with other results in the literature with steeper IMF slopes. Most notably, the \citetalias{weisz_high-mass_2015} $\overline{\Gamma}$ value of $1.45^{0.03}_{-0.06}$ for M31, but also the Kennicutt IMF ($\Gamma$ = 1.5; \citet{kennicutt_rate_1983}), the \citeauthor{miller_initial_1979} IMF from 1-10M$_{\odot}$ and the high-mass Scalo IMF ($\Gamma$ = 1.6; \citet{scalo_stellar_1986}) above 4M$_{\odot}$. This potentially steeper IMF could have drastic ramifications in the number of high-mass stars produced, which would have far reaching impacts to nearly every aspect of astronomy. Due to these high impacts, in this section, we will explore some of these implication a steeper IMF slope could have. 

Because of the homogeneity and high-precision of the measurement in M31, \citet{weisz_high-mass_2015}, recommend the community adopt the modified Kroupa IMF which takes the form,
\begin{equation}
    \xi(m) = cm^{-(\Gamma + 1)} 
    \begin{cases}
    \Gamma = 0.3 
    \hspace{32pt} \rm{for} \hspace{4pt} 0.08 < m < 0.5 M_{\odot} \\
    \Gamma = 1.3 
    \hspace{32pt} \rm{for} \hspace{4pt} 0.5 < m < 1.0 M_{\odot} \\
    \Gamma = 1.45^{+0.03}_{-0.06} 
    \hspace{10pt} \rm{for} \hspace{4pt} 1.0 < m < 100 M_{\odot} \\
    \end{cases}
\end{equation}

This modified IMF is extrapolated from 2 M$_{\odot}$ down to 1M$_{\odot}$, and up to 100 M$_{\odot}$, allowing for a fair comparison to conventional definitions in the literature with effects less than a few percent. Assuming this modified Kroupa IMF, we replicate the calculations from \citetalias{weisz_high-mass_2015} of number of stars expected relative the the traditional \citeauthor{kroupa_variation_2001} IMF. Because the IMF measured here in M33 is statistically consistent with that of M31, the implications are the same as Section 5.3 of \citetalias{weisz_high-mass_2015}. Here, we highlight the largest impacts from \citetalias{weisz_high-mass_2015}, and include new estimates for astronomical fields which have become more refined in the time since \citetalias{weisz_high-mass_2015}. 

Most notably, the modified Kroupa IMF predicts the formation of fewer stars than \citeauthor{kroupa_variation_2001} at all masses, and fewer high-mass stars than a \citeauthor{salpeter_luminosity_1955} IMF for masses $>2$M$_{\odot}$. For the most massive stars, the modified Kroupa model predicts $\sim25\%$ fewer stars with masses above 8M$_{\odot}$ compared to a \citeauthor{kroupa_variation_2001} IMF. The smaller number of these most massive stars has the largest impacts for the field, as the 8M$_{\odot}$ stars are believed to be the progenitors for core collapse supernovae \citep[e.g.,][]{smartt_progenitors_2009}. These supernovae serve as a primary mechanism for alpha enhancement, and thus our ability to correctly model chemical enrichment \citep[e.g.,][]{buck_challenge_2021}. 

These stars with masses greater than 8M$_{\odot}$ are believed to be the progenitors for neutron stars and stellar mass black holes \citep[e.g.,][]{heger_how_2003}, and are the progenitors for the newly detected gravitational waves \citep[e.g.,][]{abbott_gw170817_2017, abbott_gw190521_2020}. The number of available progenitors decreasing by even a factor of a few will have drastic impacts on the inferred merger-rates \citep{abbott_population_2023}, and the number of expected gravitational wave detections, impacting the open questions such as the black hole mass gap \citep{farmer_mind_2019}. 

On a similar thread, fewer massive stars will decrease the number of expected Gamma-Ray-Bursts (GRBs) \citep{heger_how_2003}. GRBs have been used to infer star formation histories at both high and low redshifts \citep[e.g,][]{taggart_core-collapse_2021, wang_evolving_2011}. Adapting the \citet{wang_evolving_2011} GRB rates, $25\%$ fewer expected GRBs will under-predict the inferred star formation histories in high z galaxies by as much as a factor of $\sim1.2$. This additional steepness of the IMF also requires more than the factor of 0.77 correction on the cosmic star formation history scale factor of \citet{hopkins_normalization_2006}, providing additional evidence of an evolving IMF throughout cosmic time, where the number of massive stars created today is different than the earliest generations of stars. The number of high-mass stars is also imperative to help answer some of the outstanding questions on whether massive stars form through direct collapse or mergers \citep[e.g,][]{bally_birth_2005}.

\citetalias{weisz_high-mass_2015} also discuss the implications for common SFR indicators like the luminosity to SFR relation, and H$\alpha$ tracers \citep{kennicutt_star_1998, conroy_propagation_2009,conroy_propagation_2010, kennicutt_star_2012}. Specifically, for a fixed luminosity, the SFRs assuming the modified Kroupa IMF are a factor of $\sim1.3-1.5$ larger. 

Another possible explanation for the steeper IMF slope could be that observational measurements of cluster stars have been impacted by high-mass stars being preferentially ejected from clusters within the first few Myr of a clusters life. \citet{oh_dynamical_2016} have shown that the \citetalias{weisz_high-mass_2015} slope could be consistent with such a model, but we cannot distinguish this possibility from a steeper IMF with the present study. It would be of interest to look for additional observational tracers that high-mass stars that have been ejected in simulations, and see if that evidence is present in the data. 

\section{Summary} \label{sec: Summary}

We measure the high-mass IMF in Local Group galaxy M33 using data from the PHATTER survey \citep{williams_panchromatic_2021}. For the purest sample of 9 young stellar clusters identified by \citet{johnson_panchromatic_2022}, the best fit IMF is inferred to be $\overline{\Gamma} = 1.49\pm0.18$, formally consistent with the canonical \citeauthor{kroupa_variation_2001} and \citeauthor{salpeter_luminosity_1955} values. We do not find a robust trend of MF slope on the cluster's age, mass, cluster radius, galactic radius, or SFR around the cluster location. 

The results presented here are in agreement with that of M31 \citepalias{weisz_high-mass_2015}. If this increased steepness is confirmed, it has drastic ramifications in the number of high-mass stars produced and will have far reaching impacts to nearly every aspect of astronomy. 

\vspace{5mm}


\begin{acknowledgments}
Support for this work was provided by NASA through grant \#GO-14610 from the Space Telescope Science Institute, which is operated by the Association of Universities for Research in Astronomy, Incorporated, under NASA contract NAS5-26555.We recognize and thank the $\sim$2,800 Local Group Cluster Search volunteers who made this work possible. Their contributions are acknowledged individually at \url{http://authors.clustersearch.org}. This work fulfilled the requirements for TWs qualify exam, and received input from the UW qual committee, with special thanks going to James Davenport, Scott Anderson, Jess Werk, Vikki Meadows, and Zeljko Ivezic. TW and BFW were partially supported by GO-14610. The Flatiron Institute is funded by the Simons Foundation. ER acknowledges the support of the Natural Sciences and Engineering Research Council of Canada (NSERC), funding reference number RGPIN-2022-03499.
\end{acknowledgments}

\facilities{HST(ACS, WFC3)}
\software{emcee \citep{foreman-mackey_emcee_2013}, corner \citep{foreman-mackey_cornerpy_2016}, astropy \citep{astropy_collaboration_astropy_2013, astropy_collaboration_astropy_2018, astropy_collaboration_astropy_2022}, numpy \citep{harris_array_2020}, scipy \citep{virtanen_scipy_2020}, matplotlib \citep{hunter_matplotlib_2007}}


\bibliographystyle{aasjournal}
\bibliography{references.bib}

\end{document}